\documentclass[twocolumn,showpacs,preprintnumbers,amsmath,amssymb]{revtex4}
\usepackage{graphicx}% Include figure files
\usepackage{dcolumn}% Align table columns on decimal point
\usepackage{bm}% bold math

\begin{document}

\preprint{APS/123-QED}

\title{Multimode analysis of the light emitted from a
pulsed optical parametric oscillator}

\author{Anne E. B. Nielsen and Klaus M{\o}lmer}
\affiliation{Lundbeck Foundation Theoretical Center for Quantum
System Research, Department of Physics and Astronomy, University of
Aarhus, DK-8000 \AA rhus C, Denmark}

\date{\today}

\begin{abstract}
We present a multimode treatment of the optical parametric
oscillator, which is valid for both pulsed and continuous-wave pump
fields. The two-time correlation functions of the output field are
derived, and we apply the theory to analyze a scheme for heralded
production of non-classical field states that may be subsequently
stored in an atomic quantum memory.
\end{abstract}

\pacs{42.50.Dv, 03.65.Wj, 03.67.-a}% PACS, the Physics and Astronomy
                             % Classification Scheme.
%\keywords{Suggested keywords}%Use showkeys class option if keyword
                              %display desired

\maketitle

\section{Introduction}

The process of parametric down conversion, in which pump photons
with frequency $\omega$ are sent through a nonlinear crystal and
converted into pairs of signal and idler photons with frequencies
$\omega_1$ and $\omega_2=\omega-\omega_1$, is a well-known and
widely used phenomenon in quantum optics. Since the conversion
efficiency is rather small, in the optical parametric oscillator
(OPO) the process is enhanced by placing the crystal inside an
optical cavity. Due to its importance as a source of squeezed light
when operated below threshold, the OPO has been subject of much
investigation, in particular for the case of a time independent
continuous-wave pump field \cite{collett,drummond,lu}. The case of a
periodically modulated pump field has been studied in
\cite{adamyan}. In the standard treatment of the OPO, the
calculations are performed in frequency space, the pump is taken as
a time independent monochromatic beam, and it is often assumed that
the cavity allows only one or two modes with well-defined
frequencies, depending on whether degenerate or nondegenerate
operation is considered. The field leaking out of the cavity is then
determined from input-output formalism.

An OPO may also be pumped with a pulsed light field. This case is
often discussed as if the OPO generates a single mode pulse of
squeezed light, but this is not quite the case. The output consists
of a number of independent modes squeezed by different amounts as
discussed for single pass down conversion in Refs.\
\cite{bennink,wasilewski}. In the present paper we use a completely
different approach to characterize the output from a pulsed OPO. The
treatment is a generalization of the discussion of the
continuous-wave OPO in Ref.\ \cite{petersen}, and in contrast to
Refs.\ \cite{collett,drummond,lu,bennink,wasilewski} all
calculations are performed in the time domain.

The time domain treatment has several advantages. Firstly, it is not
necessary to assume the existence of a single mode or a couple of
independent modes in the cavity since all effects of cavity
resonances appear naturally in the analysis from the specification
of the length of the cavity via the cavity round trip time.
Furthermore, since we do not assume orthogonal cavity modes, our
treatment is valid for all values of the transmission of the OPO
output coupling mirror and not just for small transmissions. In
particular, we may increase the mirror transmission to unity in our
formulas and obtain the results for single pass down conversion.
Secondly, we do not assume any particular temporal shape of the pump
field, and the analysis is thus suitable to investigate the
transition between the few mode situation for short pulses and the
highly multimode regime for continuous-wave fields. In the limit of
a time independent pump field our results reduce to those of Ref.\
\cite{collett}, provided we invoke extra approximations which
guarantee the existence of only a single cavity mode. Thirdly, the
time domain analysis is convenient if the light emitted from the OPO
is detected continuously in time and the back action of the
measurements on the system has to be taken into account, see
\cite{petersen}.

\begin{figure}
\begin{center}
\includegraphics*[viewport=50 105 280 235,width=0.75\columnwidth]{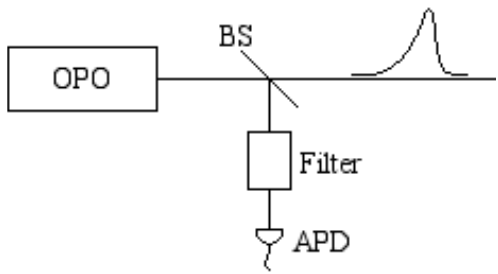}
\end{center}
\caption{Experimental setup to generate nonclassical states of
light. BS denotes a beam splitter with a small reflectance, and the
curve to the right is an example of a possible mode function that
may contain a non-Gaussian state conditioned on a detector click in
the APD.} \label{setup}
\end{figure}

Although the OPO output can be both squeezed and entangled, it
belongs to the family of so-called Gaussian states, and for a number
of applications in quantum communication and computing, non-Gaussian
states are necessary. The OPO output can be transformed into
non-Gaussian states by measurements as shown schematically in Fig.\
\ref{setup}: a small fraction of the light is extracted, possibly
frequency filtered, and detected by an avalanche photo diode (APD).
As, e.g., a squeezed vacuum state is a superposition of even photon
number states, a photo detection in the APD heralds the generation
of a state with only odd photon number contributions such as single
photon states and odd Schr\"odinger kitten states in the remaining
beam as verified experimentally in
\cite{ourjoumtsevcat,neergaardcat,wakui}. Ref.\ \cite{nielsen1}
provides a theoretical determination of the mode of the output state
with the largest single photon fidelity for the case of a
continuous-wave nondegenerate OPO. We note that in the experimental
verification by homodyne detection, one may use a constant amplitude
local oscillator, measure the homodyne detector signal as a function
of time, and then at a later stage multiply the recorded data with
the mode function, selected according to the instant of the photo
detection by the APD \cite{neergaardcat}.

For several applications of the heralded non-Gaussian field states
it is, however, necessary to mode match the state in real time. This
is the case if one wants to interfere different pulses and if one
wants to store the field state in an atomic medium and, e.g.,
perform quantum gates by optical non-linearities in the medium
\cite{andre}. Whether making use of electromagnetically induced
transparency \cite{eit}, a Raman transfer \cite{storeraman}, or
Faraday polarization rotation \cite{storerot,sherson}, storage of
light pulses in atomic media involves a strong pump pulse that
defines the temporal mode of the quantized field to be mapped onto
the atomic collective degree of freedom \cite{storegen}. This
implies that storage of a field state heralded by an APD detection
event requires a storage control field, which has to be produced at
the appropriate time. Since the mode function occupied by the
non-Gaussian state may extend to times both before and after the
trigger event \cite{nielsen1}, it may thus be necessary to switch on
the control field \textit{before} the trigger event or, more
realistically, to delay the arrival of the signal to the storage
medium. For demonstration experiments, it would be attractive to
apply the storage protocol irrespective of the APD output and
subsequently post select the instances where APD detection actually
occurred in a suitable time window around the instant for which the
stored mode is optimal. In this paper we do not model the storage
process itself, but we identify the mode functions that maximize the
non-Gaussian features of the pulses to be stored, conditioned on
realistic detection events. In particular, we wish to identify mode
functions that are optimal independent of the precise detection time
within a predefined time interval, which can not be chosen
arbitrarily small if one wants a satisfactory APD detection
probability. We shall for concreteness take the optimal mode to be
the one with the most negative value of the Wigner function at the
origin.

The paper is structured as follows. In Sec.\ II we develop a
theoretical description of the OPO and derive the two-time
correlation functions of the OPO output field. In Sec.\ III we show
that a treatment similar to the one applied to the OPO can be used
to express the output field from the filter in Fig.\ \ref{setup} in
terms of the input field. In Sec.\ IV we determine the mode of the
conditioned state with the most negative value of the Wigner
function at the origin, we compute the probability to obtain a
trigger detection event within a specified time interval, and we
compare the performance of continuous-wave and pulsed operation of
the OPO for preparation of pulses that can be stored with predefined
control pulses in an atomic medium. Section V concludes the paper.

\section{Correlation functions for the optical parametric oscillator output field}

\begin{figure}
\begin{center}
\includegraphics*[viewport=23 20 310 240,width=0.75\columnwidth]{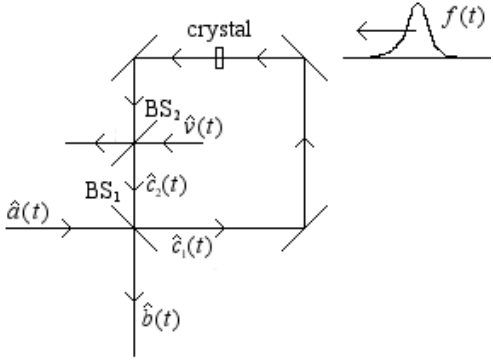}
\end{center}
\caption{Model of an optical parametric oscillator. Photon pairs are
generated by parametric down conversion in the crystal, which is
pumped by the classical field $f(t)$. The beam splitter
$\textrm{BS}_1$ couples the intra cavity field to the input (vacuum)
field $\hat{a}(t)$ and the output field $\hat{b}(t)$, and the beam
splitter $\textrm{BS}_2$ is included to model intra cavity losses.}
\label{OPO}
\end{figure}

A theoretical model of the OPO is based on the setup illustrated in
Fig.\ \ref{OPO}. We have chosen a four-sided ring cavity for
mathematical convenience, but this is not essential to the analysis.
The field annihilation operators of the input and the output fields
are denoted $\hat{a}(t)$ and $\hat{b}(t)$, respectively, while
$\hat{c}_i(t)$, $i=1,2$, represents the field at different positions
inside the cavity as shown, and $\hat{v}(t)$ is the annihilation
operator of a field in the vacuum state. The beam splitter
$\textrm{BS}_1$ couples the input and output fields to the intra
cavity field, and the fictitious beam splitter $\textrm{BS}_2$
models losses in the system. We model the entire loss as if it takes
place between the crystal in the upper part of the figure and the
output coupling mirror, which represents a worst case situation.

The crystal is pumped by a classical pump field
$f(t)=|f(t)|e^{i\phi_f(t)}$, which is assumed to pass unhindered
through the cavity mirrors. The parametric process in a non-linear
crystal is described in \cite{bennink,wasilewski}, and leads, in the
time domain, to the following mapping of the field incident on the
crystal to the field leaving the crystal
\begin{equation}\label{sqtrans}
\hat{c}(t)\rightarrow\cosh(2\chi|f(t)|)\hat{c}(t) -ie^{i\phi_f(t)}
\sinh(2\chi|f(t)|)\hat{c}^\dag(t).
\end{equation}
In \eqref{sqtrans} $\chi$, which is taken to be real, is
proportional to the second order susceptibility of the crystal and
to the length of the crystal. We have assumed that the bandwidth of
the down conversion is infinite and neglected depletion of the pump
field and differences in the phase matching conditions for down
conversion to different frequencies. This is likely to be a good
approximation in the experiment described in Ref.\
\cite{neergaardcat}, since the crystal inside the OPO is only few
millimeters long, and thousands of cavity modes are populated. Also,
modes far from the center of the spectrum are irrelevant because
they are filtered out by other components in experiments (see the
next section).

In order to express $\hat{b}(t)$ in terms of $\hat{a}(t)$ and
$\hat{v}(t)$, we use the beam splitter relations
\begin{eqnarray}
\hat{b}(t)&=&t_1\hat{c}_2(t)+ir_1\hat{a}(t)\label{BSa}\\
\hat{c}_1(t)&=&ir_1\hat{c}_2(t)+t_1\hat{a}(t)\label{BSb}
\end{eqnarray}
and express $\hat{c}_2(t)$ in terms of $\hat{c}_1(t-\tau)$
\begin{multline}\label{c2}
\hat{c}_2(t)=-it_2\big(\cosh\left(z(t-\tau)\right)\hat{c}_1(t-\tau)\\
-ie^{i\phi(t-\tau)}\sinh\left(z(t-\tau)\right)
\hat{c}_1^\dag(t-\tau)\big)+ir_2\hat{v}(t).
\end{multline}
$t_i$ ($r_i$) is the field transmission (reflection) coefficient of
$\textrm{BS}_i$, and $\tau$ is the round trip time in the cavity. In
\eqref{c2}, we have assumed that each cavity mirror induces a phase
shift of $\pi/2$, and we have defined the following quantities
\begin{eqnarray}
z(t-\tau)&\equiv&2\chi|f(t-\tau+\xi\tau)|\\
\phi(t-\tau)&\equiv&\phi_f(t-\tau+\xi\tau),
\end{eqnarray}
where $\xi\tau$ is the traveling time from $\textrm{BS}_1$ to the
crystal. Furthermore, for convenience, we have redefined
$\hat{v}(t)$ according to
$\hat{v}(t-\zeta\tau)\rightarrow\hat{v}(t)$, where $\zeta\tau$ is
the traveling time from $\textrm{BS}_2$ to $\textrm{BS}_1$.
Isolating $\hat{b}(t)$ from Eqs.\ \eqref{BSa}, \eqref{BSb}, and
\eqref{c2} and assuming $e^{i\phi(t)}=e^{i\phi(t-\tau)}$ for all
$t$, we find
\begin{multline}\label{bt}
\hat{b}(t)=ir_1\hat{a}(t)-it_1^2\sum_{n=0}^\infty r_1^nt_2^{n+1}\\
\Bigg(\cosh\left(\sum_{k=1}^{n+1}z(t-k\tau)\right)\hat{a}(t-(n+1)\tau)\\
-ie^{i\phi(t)}\sinh\left(\sum_{k=1}^{n+1}z(t-k\tau)\right)
\hat{a}^\dag(t-(n+1)\tau)\Bigg)\\
+it_1r_2\hat{v}(t)+it_1r_2
\sum_{n=0}^\infty (r_1t_2)^{n+1}\\
\Bigg(\cosh\left(\sum_{k=1}^{n+1}z(t-k\tau)\right)\hat{v}(t-(n+1)\tau)\\
-ie^{i\phi(t)}\sinh\left(\sum_{k=1}^{n+1}z(t-k\tau)\right)
\hat{v}^\dag(t-(n+1)\tau)\Bigg).
\end{multline}
The requirement $e^{i\phi(t)}=e^{i\phi(t-\tau)}$, which is satisfied
for $\phi(t)=-2\pi Nt/\tau+\phi_0$, where $N$ is an integer and
$\phi_0$ is a constant, means that the successive squeezing
operations add up in phase, i.e., it is the same quadrature that is
squeezed at times $t$, $t\pm\tau$, $t\pm2\tau$, $\ldots$ .

For the special case of a time independent continuous-wave pump
field $z(t)=z$, below threshold $r_1t_2e^{z}<1$, a Fourier transform
of \eqref{bt} leads to the following expression for the output field
in frequency domain:
\begin{multline}\label{26}
\hat{b}(\omega_0+\omega)=G_1(\omega+\omega_0)\hat{a}(\omega_0+\omega)+
g_1(\omega-\omega_0)\hat{a}^\dag(\omega_0-\omega)\\
+G_2(\omega+\omega_0)\hat{v}(\omega_0+\omega)
+g_2(\omega-\omega_0)\hat{v}^\dag(\omega_0-\omega),
\end{multline}
where $\omega_0=N\pi/\tau$ is half the frequency of the pump field,
\begin{eqnarray}
G_1(\omega+\omega_0)&=&ir_1-i\frac{t_1^2t_2}{2}G_+(\omega+\omega_0),\\
G_2(\omega+\omega_0)&=&it_1r_2+i\frac{t_1r_1t_2r_2}{2}G_+(\omega+\omega_0),\\
g_1(\omega-\omega_0)&=&-e^{i\phi_0}\frac{t_1^2t_2}{2}G_-(\omega-\omega_0),\\
g_2(\omega-\omega_0)&=&e^{i\phi_0}\frac{t_1r_1t_2r_2}{2}G_-(\omega-\omega_0),
\end{eqnarray}
and
\begin{equation}
G_\pm(\omega)=\frac{e^{z+i\omega\tau}}{1-r_1t_2e^{z+i\omega\tau}}\pm
\frac{e^{-z+i\omega\tau}}{1-r_1t_2e^{-z+i\omega\tau}}.
\end{equation}
Assuming that the input field $\hat{a}(t)$ is in the vacuum state,
the frequency correlation function
\begin{multline}
\langle\hat{b}^\dag(\omega)\hat{b}(\omega')\rangle=
t_1^2t_2^2(1-r_1^2t_2^2)\sinh^2(z)\delta(\omega-\omega')\big/\\
\Big(1+r_1^4t_2^4+4r_1^2t_2^2\cosh^2(z)+2r_1t_2
\big(r_1t_2\cos(2\omega\tau)\\
-2(1+r_1^2t_2^2)\cosh(z)\cos(\omega\tau)\big)\Big)
\end{multline}
shows that resonances occur for $\omega\tau=n2\pi$, $n\in Z$, as
expected. In particular, the degenerate frequency $\omega_0$ is only
a resonance frequency if $N$ is even. To compare Eq.\ \eqref{26} to
the results given in Ref.\ \cite{collett} for a cavity with a single
resonance frequency at $\omega_0$, we assume $N$ to be even and
$\omega\tau\ll1$ in Eq.\ \eqref{26}. In order to have well separated
cavity modes we must also assume $t_1^2\ll1$ and $r_2^2\ll1$, but to
maintain a finite width of the modes as in Ref.\ \cite{collett} we
keep $\gamma_1\equiv t_1^2/\tau$ and $\gamma_2\equiv r_2^2/\tau$
fixed. Since $\tau\gamma_1\ll1$ and $\tau\gamma_2\ll1$ in this
limit, the mean number of round trips in the cavity is large, and
$z$ must be correspondingly small, i.e., $z$ must be proportional to
$\tau$. For $|z|=|\epsilon|\tau$ we obtain Eq.\ 46 of Ref.\
\cite{collett}.

Returning to the general case, we compute the two-time correlation
functions of the output field for $\hat{a}(t)$ in the vacuum state
from Eq.\ \eqref{bt}:
\begin{multline}\label{bdb}
\langle\hat{b}^\dag(t)\hat{b}(t')\rangle=t_1^2t_2^2(1-r_1^2t_2^2)\\
\Bigg(\sum_{q=1}^\infty\sum_{m=0}^\infty(r_1t_2)^{q+2m}
\sinh\left(\sum_{k=1}^{m+1}z(t-k\tau)\right)\\
\sinh\left(\sum_{k=1}^{q+m+1}z(t'-k\tau)\right)\delta(t-t'+q\tau)+\\
\sum_{q=0}^\infty\sum_{m=0}^\infty(r_1t_2)^{q+2m}
\sinh\left(\sum_{k=1}^{q+m+1}z(t-k\tau)\right)\\
\sinh\left(\sum_{k=1}^{m+1}z(t'-k\tau)\right)\delta(t-t'-q\tau)\Bigg)
\end{multline}
and
\begin{multline}\label{bb}
\langle\hat{b}(t)\hat{b}(t')\rangle=ie^{i\phi(t)}t_1^2t_2^2(1-r_1^2t_2^2)\\
\Bigg(\sum_{q=1}^\infty\sum_{m=0}^\infty(r_1t_2)^{q+2m}
\sinh\left(\sum_{k=1}^{m+1}z(t-k\tau)\right)\\
\cosh\left(\sum_{k=1}^{q+m+1}z(t'-k\tau)\right)\delta(t-t'+q\tau)+\\
\sum_{q=0}^\infty\sum_{m=0}^\infty(r_1t_2)^{q+2m}
\cosh\left(\sum_{k=1}^{q+m+1}z(t-k\tau)\right)\\
\sinh\left(\sum_{k=1}^{m+1}z(t'-k\tau)\right)\delta(t-t'-q\tau)\Bigg).
\end{multline}
These two expressions and $\langle\hat{b}(t)\rangle=0$ are
sufficient to characterize the output state completely, because it
is Gaussian.

The correlations of an arbitrary single mode with mode function
$h(t)$ and annihilation operator
\begin{equation}\label{singlemode}
\hat{b}=\int h^*(t)\hat{b}(t)dt
\end{equation}
are easily calculated from Eqs.\ \eqref{bdb} and \eqref{bb}. We
shall only be concerned with the degenerate case below, and we thus
choose
$h(t)=\tilde{h}(t)e^{i\phi_0/2+i\pi/4}e^{-i\omega_0t}=\tilde{h}(t)e^{i\phi(t)/2+i\pi/4}$
with $\tilde{h}(t)$ real. The constant phase factor is included to
obtain a real value of $\langle\hat{b}^2\rangle$, which corresponds
to the case, where the axes of the squeezing ellipse in phase space
lie along the quadrature axes. In the following we take $h(t)$ to be
real and omit the factor $ie^{i\phi(t)}$ in Eq.\ \eqref{bb}.

The variances of $\hat{x}\equiv(\hat{b}+\hat{b}^\dag)/\sqrt{2}$ and
$\hat{p}\equiv-i(\hat{b}-\hat{b}^\dag)/\sqrt{2}$ are
$1/2+\langle\hat{b}^\dag\hat{b}\rangle+\langle\hat{b}^2\rangle$ and
$1/2+\langle\hat{b}^\dag\hat{b}\rangle-\langle\hat{b}^2\rangle$,
respectively, and the most efficiently squeezed mode is thus
obtained by choosing $h(t)$ as the eigenfunction with the smallest
eigenvalue $\lambda$ of an integral equation with either the
integration kernel
$\delta(t-t')/2+\langle\hat{b}^\dag(t)\hat{b}(t')\rangle+
\langle\hat{b}(t)\hat{b}(t')\rangle$ or
$\delta(t-t')/2+\langle\hat{b}^\dag(t)\hat{b}(t')\rangle-
\langle\hat{b}(t)\hat{b}(t')\rangle$, and $\lambda$ is the
corresponding variance. Since the output field at time $t$ is only
correlated to the output field at times $t+n\tau$, $n\in Z$, the
mode function of the most efficiently squeezed mode is in general a
spike function that is only nonzero at times $t+n\tau$,
$n=\ldots,-1,0,1,2,\ldots$ , for some $t$, but if the variations of
the pump field take place on a time scale $\tau$ or more slowly,
displaced spike functions, and therefore also a smooth mode function
with the same envelope, will be only slightly less squeezed.

In the limit of a time independent pump field, Eqs.\ \eqref{bdb} and
\eqref{bb} reduce to
\begin{multline}
\langle\hat{b}^\dag(t)\hat{b}(t')\rangle=
\frac{t_1^2t_2^2}{4}\sum_{q=-\infty}^{\infty}(r_1t_2)^{|q|}\\
\bigg(\frac{1-r_1^2t_2^2}{1-r_1^2t_2^2e^{2z}}e^{(2+|q|)z}+
\frac{1-r_1^2t_2^2}{1-r_1^2t_2^2e^{-2z}}e^{-(2+|q|)z}\\
-e^{|q|z}-e^{-|q|z}\bigg)\delta(t-t'-q\tau)
\end{multline}
and
\begin{multline}
\langle\hat{b}(t)\hat{b}(t')\rangle=ie^{i\phi(t)}\frac{t_1^2t_2^2}{4}
\sum_{q=-\infty}^{\infty}(r_1t_2)^{|q|}\\
\bigg(\frac{1-r_1^2t_2^2}{1-r_1^2t_2^2e^{2z}}e^{(2+|q|)z}-
\frac{1-r_1^2t_2^2}{1-r_1^2t_2^2e^{-2z}}e^{-(2+|q|)z}\\
-e^{|q|z}+e^{-|q|z}\bigg)\delta(t-t'-q\tau).
\end{multline}

Until now we have described the fields in terms of time dependent
Heisenberg picture operators, but it is also useful to consider the
OPO model from a Schr\"odinger picture point of view. In particular,
the latter approach is suitable for a numerical treatment of the
pulsed OPO. To this end we divide the light beams into small
segments of (infinitesimal) duration $\Delta t$ and treat each
segment as a single mode, i.e., we define
\begin{equation}
g_i(t)=\left\{\begin{array}{cl} 1/\sqrt{\Delta t}&\textrm{for
}t_i-\Delta t/2\leq t<t_i+\Delta t/2\\
0&\textrm{otherwise}
\end{array}\right.
\end{equation}
and replace the continuous field annihilation operator $\hat{d}(t)$
($\hat{d}(t)=\hat{a}(t)$, $\hat{b}(t)$, $\hat{c}_1(t)$, or
$\hat{c}_2(t)$) by the discrete annihilation operators
\begin{equation}\label{disope}
\hat{d}_i=\int g_i^*(t)\hat{d}(t)dt=\hat{d}(t_i)\sqrt{\Delta t}
\end{equation}
localized at time $t_i$. In practice, we will have to deal with a
finite $\Delta t$, and neglecting the field variation within each
$\Delta t$ interval is equivalent to a cutoff in frequency,
justified by the experimentally relevant frequency regime. A
thorough discussion of continuous and discrete operator descriptions
of light beams in both time and frequency space may be found in
Ref.\ \cite{blow}. Since the output field and the cavity field are
Gaussian for a vacuum input field, the state of all the small light
beam segments is efficiently represented by a Wigner function. In
general, the Wigner function of an $n$-mode Gaussian state with zero
mean values is on the form
\begin{equation}\label{wigner}
W(y)=\frac{1}{\pi^n\sqrt{\det(V)}}e^{-y^TV^{-1}y},
\end{equation}
where $y=(x_1,p_1,x_2,p_2,\ldots,x_n,p_n)^T$ is a column vector of
quadrature variables, and $V$ is the covariance matrix of the $n$
modes.
$V=\langle\hat{y}\hat{y}^T\rangle+\langle\hat{y}\hat{y}^T\rangle^T$,
where
$\hat{y}=(\hat{x}_1,\hat{p}_1,\hat{x}_2,\hat{p}_2,\ldots,\hat{x}_n,\hat{p}_n)^T$,
$\hat{x}_i=(\hat{d}_i+\hat{d}^\dag_i)/\sqrt{2}$,
$\hat{p}_i=-i(\hat{d}_i-\hat{d}^\dag_i)/\sqrt{2}$, and $\hat{d}_i$
is the field annihilation operator of mode $i$. Before the pump
pulse reaches the crystal, all modes are in the vacuum state, and
$V$ is the identity matrix. We include all the $\tau/\Delta t$
cavity modes and a sufficiently large number of $\hat{a}(t)$ and
$\hat{v}(t)$ modes in \eqref{wigner}. As time passes by modes are
squeezed, when they hit the crystal, input and cavity modes are
transformed into cavity and output modes, when they hit
$\textrm{BS}_1$, and vacuum and cavity modes are transformed into
cavity and lost modes, when they hit $\textrm{BS}_2$. For each of
these transformations $\hat{y}$ is transformed according to
$\hat{y}\rightarrow S\hat{y}$, where $S$ is a matrix, which is
easily determined from Eqs.\ \eqref{BSa}, \eqref{BSb}, and
\eqref{c2}. The corresponding transformation of $V$ is $V\rightarrow
SVS^T$. At the end of the calculation all rows and columns of $V$
that represent lost modes are erased, and the result is a matrix,
which contains the same information as Eqs.\ \eqref{bdb} and
\eqref{bb}. We have used this method to compute the results for the
pulsed OPO in Sec.\ IV.

\section{Filtering}

\begin{figure}
\begin{center}
\includegraphics*[viewport=20 30 290 220,width=0.75\columnwidth]{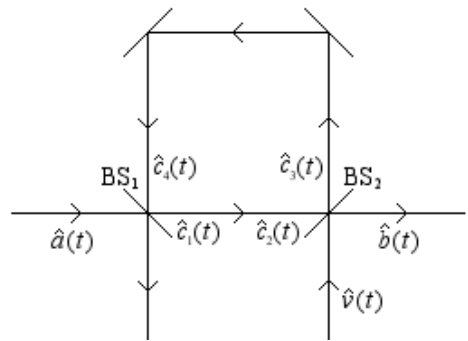}
\end{center}
\caption{Filter cavity. The input field $\hat{a}(t)$ is partially
transmitted into the cavity through the beam splitter
$\textrm{BS}_1$. The reflected component of the input field is lost
as is the part of the intra cavity field that leaves the cavity
through $\textrm{BS}_1$. The part of the intra cavity field that
leaves the cavity through $\textrm{BS}_2$ contributes to the output
field $\hat{b}(t)$. $\hat{v}(t)$ is a field in the vacuum state.}
\label{filter}
\end{figure}

The OPO produces pairwise quantum correlated fields in a large
number of cavity field modes, and in experiments it is necessary to
apply a frequency filter before the trigger detector, see Fig.\
\ref{setup}, to ensure that the photon is derived from a well
defined sideband, and that the signal field predominantly occupies a
single mode. We now turn to a description of such a filter modeled
by a cavity with two beam splitters $\textrm{BS}_1$ and
$\textrm{BS}_2$ and two perfectly reflecting mirrors as depicted in
Fig.\ \ref{filter}. Copying the notation from Sec.\ II, $\hat{a}(t)$
is the input field, $\hat{b}(t)$ is the output field, $\hat{v}(t)$
is a field in the vacuum state, $\hat{c}_i(t)$ is the field at
different positions inside the cavity (see the figure), $t_i$
($r_i$) is the field transmission (reflection) coefficient of
$\textrm{BS}_i$, and $\tau_{\mathrm{F}}$ is the round trip time in
the filter cavity. The fields are related according to the equations
\begin{eqnarray}
\hat{c}_1(t)&=&t_1\hat{a}(t)+ir_1\hat{c}_4(t),\\
\hat{c}_2(t)&=&\hat{c}_1(t-\tau_{\mathrm{F}}/4),\\
\hat{c}_3(t)&=&t_2\hat{v}(t)+ir_2\hat{c}_2(t),\\
\hat{c}_4(t)&=&-\hat{c}_3(t-3\tau_{\mathrm{F}}/4),\\
\hat{b}(t)&=&t_2\hat{c}_2(t)+ir_2\hat{v}(t),
\end{eqnarray}
from which we derive
\begin{multline}\label{bFt}
\hat{b}(t)=ir_2\hat{v}(t)-ir_1t_2^2
\sum_{n=0}^\infty(r_1r_2)^n\hat{v}(t-(n+1)\tau_{\mathrm{F}})\\
+t_1t_2\sum_{n=0}^\infty(r_1r_2)^n\hat{a}(t-(n+1/4)\tau_{\mathrm{F}}).
\end{multline}
Transforming the output field to frequency space
\begin{multline}\label{bFomega}
\hat{b}(\omega)=ir_2\hat{v}(\omega)
-\frac{ir_1t_2^2e^{i\omega\tau_{\mathrm{F}}}}
{1-r_1r_2e^{i\omega\tau_{\mathrm{F}}}}\hat{v}(\omega)\\
+\frac{t_1t_2e^{i\omega\tau_{\mathrm{F}}/4}}
{1-r_1r_2e^{i\omega\tau_{\mathrm{F}}}}\hat{a}(\omega),
\end{multline}
we find
\begin{equation}
\langle\hat{b}^\dag(\omega)\hat{b}(\omega)\rangle=
\frac{t_1^2t_2^2}{1+r_1^2r_2^2-2r_1r_2\cos(\omega\tau_{\mathrm{F}})}
\langle\hat{a}^\dag(\omega)\hat{a}(\omega)\rangle,
\end{equation}
and it is apparent that the frequencies transmitted most efficiently
through the filter are those which satisfy the condition
$\omega\tau_{\mathrm{F}}=2\pi n$, $n\in Z$. The free spectral range
of the filter is thus $2\pi/\tau_{\mathrm{F}}$, and for a given
$\tau_{\mathrm{F}}$ the bandwidth is determined by $t_1$ and $t_2$.

In order to select just one of the frequency modes emerging from the
OPO, the bandwidth of the filter should be much smaller than the
free spectral range of the OPO, but larger than the bandwidth of the
OPO. Also, the free spectral range of the filter should be large
compared to the bandwidth of the parametric down conversion.
Experimentally it can be difficult to build cavities that satisfy
the last condition, and one may use instead a sequence of different
filter cavities \cite{neergaardcat}. Since we assumed an infinite
bandwidth of the down conversion in Sec.\ II, we must, for
consistency, use an infinitely small filter cavity in the
theoretical treatment. We thus assume
$(\omega-\omega_0)\tau_{\mathrm{F}}\ll1$, $t_1^2\ll1$, and
$t_2^2\ll1$ for fixed $\kappa_1\equiv t_1^2/\tau_{\mathrm{F}}$ and
$\kappa_2\equiv t_2^2/\tau_{\mathrm{F}}$, where $\omega_0=2\pi
n/\tau_{\mathrm{F}}$, $n\in Z$, is a resonance frequency of the
filter cavity, and expand Eq.\ \eqref{bFomega} to lowest order:
\begin{multline}\label{bFomegag}
\hat{b}(\omega)=i\left(1-\frac{\kappa_2}
{\frac{\kappa_1}{2}+\frac{\kappa_2}{2}-i(\omega-\omega_0)}\right)\hat{v}(\omega)+\\
\frac{\sqrt{\kappa_1\kappa_2}}{\frac{\kappa_1}{2}
+\frac{\kappa_2}{2}-i(\omega-\omega_0)}\hat{a}(\omega).
\end{multline}
In this limit the filter transmission function is a Lorentzian
\begin{equation}
\langle\hat{b}^\dag(\omega)\hat{b}(\omega)\rangle=
\frac{4\kappa_1^2\kappa_2^2}{(\kappa_1+\kappa_2)^2+4(\omega-\omega_0)^2}
\langle\hat{a}^\dag(\omega)\hat{a}(\omega)\rangle.
\end{equation}
We consider a single mode $h(t)$ of the output field, and define
$h_a(t)$ and $h_v(t)$ according to
\begin{equation}\label{av}
\int h^*(t)\hat{b}(t)dt=\int h^*_a(t)\hat{a}(t)dt+\int
h^*_v(t)\hat{v}(t)dt.
\end{equation}
Since the vacuum state does not contribute to normally ordered
expectation values, we shall not need $h_v(t)$ in the following, but
from Eq.\ \eqref{bFt}
\begin{equation}\label{ha}
h_a(t)=\int_t^\infty h(t')\sqrt{\kappa_1\kappa_2}e^{i\omega_0(t'-t)-
\left(\frac{\kappa_1}{2}+\frac{\kappa_2}{2}\right)(t'-t)}dt',
\end{equation}
i.e., the action of the filter is effectively to transform the mode
function $h(t)$ into the (not properly normalized) mode function
$h_a(t)$. Note that if the time dependent part of the phase of
$h(t)$ is chosen as $e^{-i\omega_0t}$, the time dependent part of
the phase of $h_a(t)$ is also given as $e^{-i\omega_0t}$. When we
choose the resonance frequency of the filter to equal half the
carrier frequency of the OPO pump beam, we may thus continue to use
real mode functions and omit the factor $e^{i\omega_0(t'-t)}$ in
Eq.\ \eqref{ha}. For a given bandwidth $\kappa_1+\kappa_2$ of the
filter,
\begin{multline}
\int|h_a(t)|^2dt=1-\int|h_v(t)|^2dt=\\
\int|h(\omega)|^2 \frac{4\kappa_1\kappa_2}{(\kappa_1+\kappa_2)^2
+4(\omega-\omega_0)^2}d\omega
\end{multline}
is maximal for $\kappa_1=\kappa_2$, and we thus assume
$\kappa_1=\kappa_2=\kappa$ in the rest of the paper.

\section{Mode function optimization for heralded generation
of a non-Gaussian state of light}

With the necessary tools at hand we can now proceed to an analysis
of the preparation of non-Gaussian light states that can be stored
in atomic samples. Gaussian states have Wigner functions that are
positive for all arguments, and as a measure of non-Gaussian
character we shall refer to negative values of the Wigner function
occurring at the origin of phase space, e.g., for odd number states
and odd Schr\"odinger cat states. In Ref.\ \cite{nielsen1} we
derived an expression for the value of the Wigner function at the
origin of the state of an arbitrary real mode of the multimode state
generated when conditioning on a photo detection event
\begin{equation}\label{Wi}
W_i(0,0)=\frac{V_{33}V_{44}(V_{11}+V_{22}-2)-V_{33}V_{24}^2-V_{44}V_{13}^2}
{\pi(V_{33}V_{44})^{3/2}(V_{11}+V_{22}-2)}.
\end{equation}
$V_{jk}$ are the elements of the Gaussian covariance matrix of the
mode in which the APD detection takes place (quadrature variables 1
and 2) and the chosen mode of the output state (quadrature variables
3 and 4) before conditioning. $V$ is computed from the definition
given just below Eq.\ \eqref{wigner} by use of the mode functions of
the two modes, the transformation \eqref{ha} of the trigger mode
function due to the filter, Eq.\ \eqref{singlemode} (with
$\hat{b}(t)$ replaced by the relevant linear combination of
$\hat{b}(t)$ and the annihilation operator of the vacuum field
entering into the system at the beam splitter in Fig.\ \ref{setup}),
and the two-time correlation functions given in Eqs.\ \eqref{bdb}
and \eqref{bb}. We take the trigger mode function to be constant in
a time interval of duration $\Delta t_t$ positioned at time $t_i$
and zero otherwise, where $\Delta t_t$ is much shorter than all
other time scales in the system. In Appendix A we discuss the
applied detector model and the choice of trigger mode function in
more detail.

As explained in the Introduction the mode that will be stored is
determined by the shape and timing of the strong storage pulse, and
it is advantageous if one can initiate the generation of the storage
pulse before it is known whether a trigger detection event will
actually take place at the right time relative to the pulse stored.
The protocol is thus probabilistic. We consider an attempted storage
as successful if a trigger detection event takes place within a
predefined time interval of duration $T$, whose position on the time
axis is determined relative to the pump pulse, if the OPO is pulsed,
and relative to the storage pulse, if the pump field is time
independent. In practice, $T$ is at least as large as the temporal
resolution of the APD detection system, which is of order 1 ns
\cite{wakui}. Since the shape of the storage pulse is independent of
the actual trigger detection time, the quantity we optimize is the
mean value of the Wigner function at the origin
\begin{equation}\label{neg}
W(0,0)=\sum_{i=1}^{T/\Delta
t_t}P_iW_i(0,0)\bigg/\sum_{j=1}^{T/\Delta t_t}P_j,
\end{equation}
where the sum is over the $T/\Delta t_t$ trigger modes inside the
acceptance interval $T$, $P_i$ is the probability to obtain a
detection event in trigger mode $i$, which for infinitesimal $\Delta
t_t$ is equal to the expectation value of the number of photons in
the $i$th trigger mode, and $W_i(0,0)$ is the value of the Wigner
function at the origin of the chosen mode of the generated state
conditioned on a detection in trigger mode $i$ given in Eq.\
\eqref{Wi}. The total trigger probability
\begin{equation}\label{P}
P=\sum_{i=1}^{T/\Delta t_t}P_i
\end{equation}
is small, and hence we neglect the possibility to have two or more
detection events within the time interval $T$.

\begin{table}
\begin{center}
\begin{tabular}{|l|c|c|}
\hline
\multicolumn{1}{|c|}{Quantity}&Symbol&Value\\
\hline Total OPO cavity length&$L$&81 cm\\
\hline Round trip time in the OPO cavity&$\tau$&2.7 ns\\
\hline Loss in cavity&$r_2^2$&0.004\\
\hline Transmission of OPO output mirror&$t_1^2$&0.127\\
\hline Reflectance of beam splitter in Fig.\ \ref{setup}&$R$&0.05\\
\hline Filter bandwidth&$2\kappa$&$5.63\cdot10^8\textrm{ s}^{-1}$\\
\hline Trigger channel efficiency&$\eta_t$&0.07\\
\hline Signal channel efficiency&$\eta_s$&0.70\\
\hline
\end{tabular}
\end{center}
\caption{Parameters used in the numerical minimization of $W(0,0)$.
$\eta_t$ includes both propagation efficiency in the trigger channel
and the efficiency of the APD, while $\eta_s$ is the propagation
efficiency of the signal channel. The values are chosen in
accordance with the numbers given in Refs.\
\cite{wakui,neergaardcat,neergaard1}}\label{values}
\end{table}

\begin{figure}
\begin{tabular}{c}
\multicolumn{1}{l}{a)}\\
\includegraphics*[viewport=6 8 383 296,width=0.95\columnwidth]{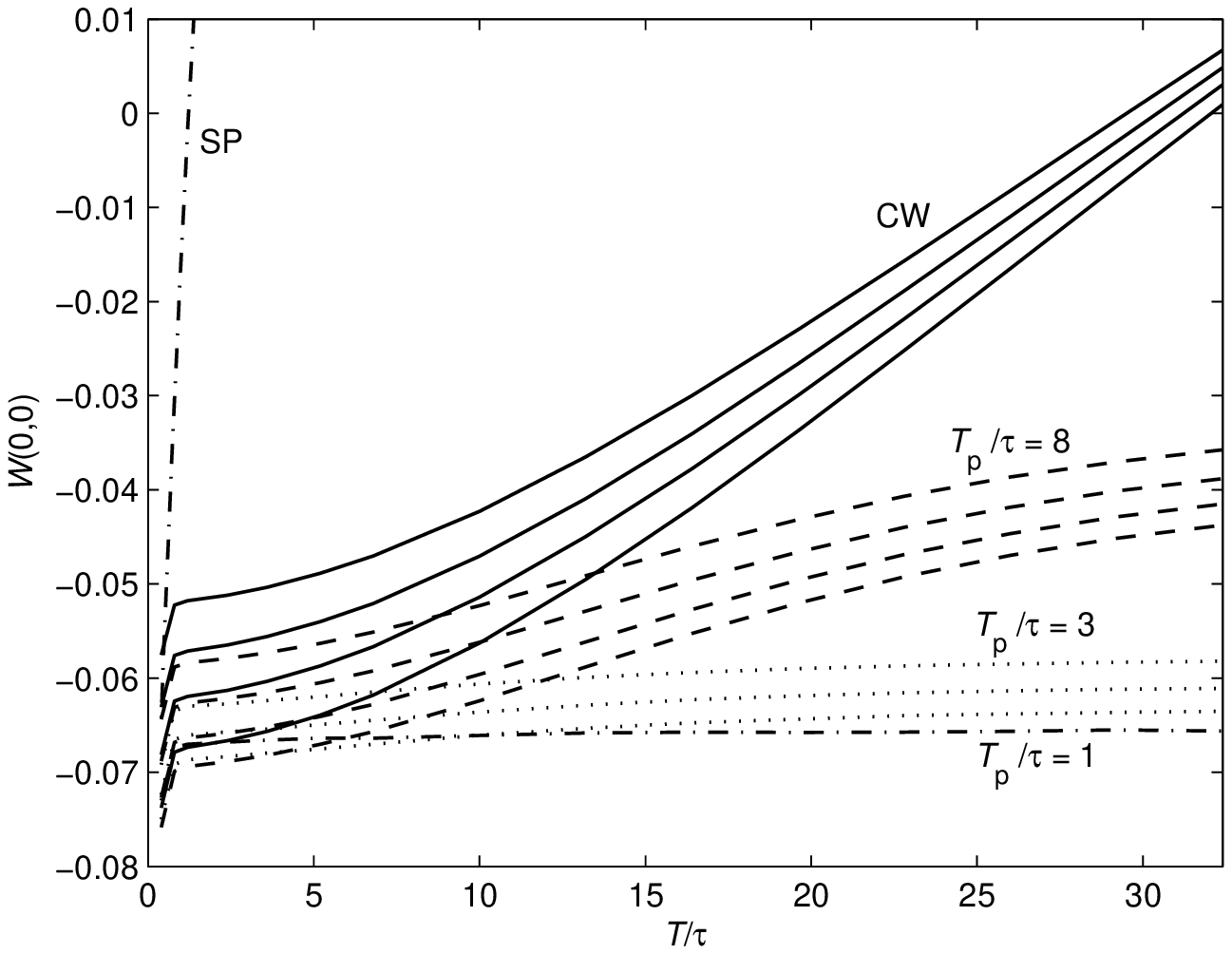}\\
\multicolumn{1}{l}{b)}\\
\includegraphics*[viewport=28 8 383 308,width=0.9\columnwidth]{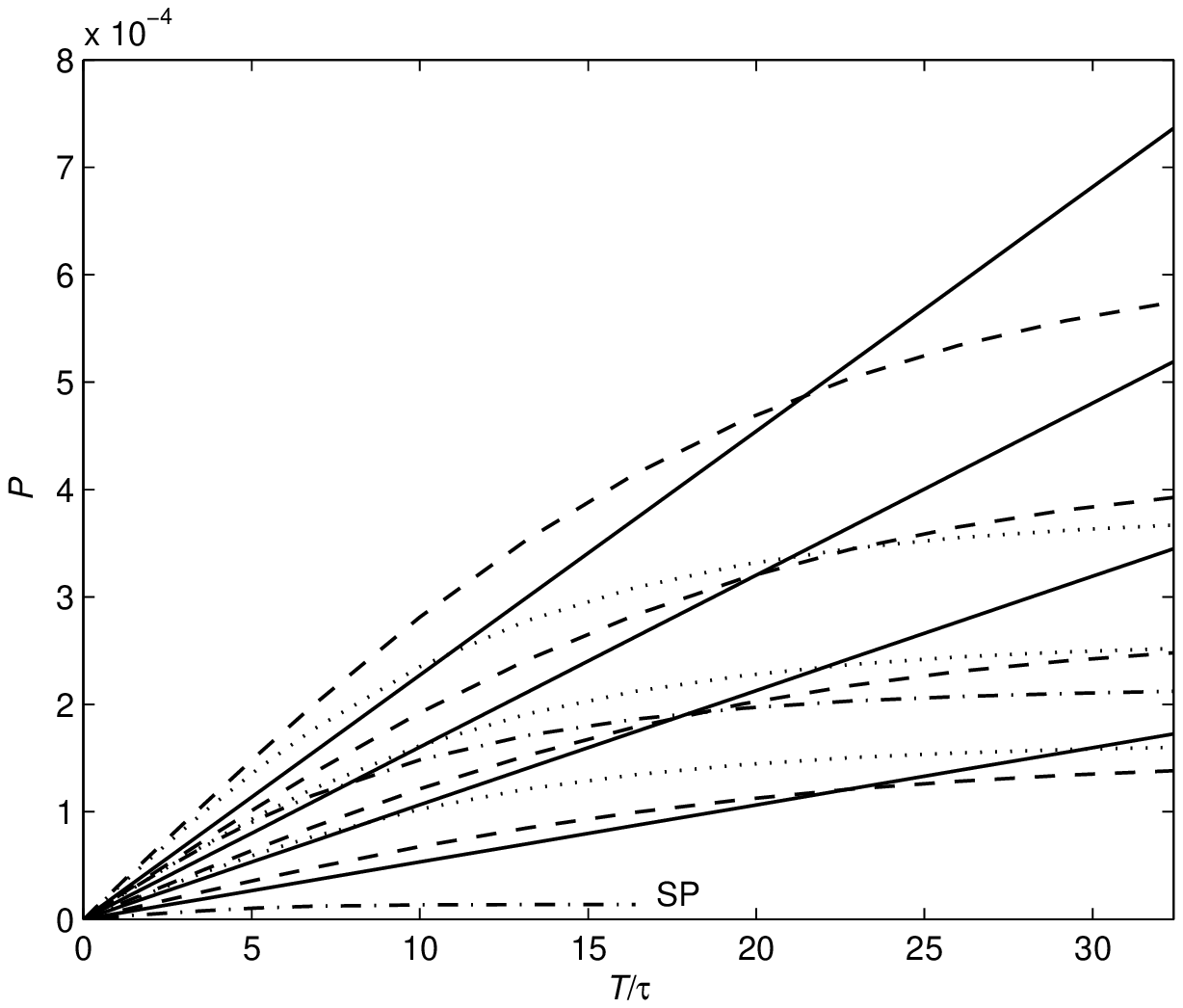}
\end{tabular}
\caption{a) $W(0,0)$ and b) success probability as a function of
$T$. Solid lines: time independent continuous-wave (CW) pump field
with $z=0.010$, $0.014$, $0.017$, and $0.020$; dashed lines: pulsed
pump field with $T_{\textrm{p}}/\tau=8$ and $s=0.03$, $0.04$,
$0.05$, and $0.06$; dotted lines: pulsed pump field with
$T_{\textrm{p}}/\tau=3$ and $s=0.04$, $0.05$, and $0.06$;
dash-dotted line: pulsed pump field with $T_{\textrm{p}}/\tau=1$ and
$s=0.07$; dash-dotted line labeled SP: single pass (SP) operation
for a pulsed pump field with $T_{\textrm{p}}/\tau=3$ and $s=0.05$.
Within each series a smaller $s$ or $z$ corresponds to a more
negative value of $W(0,0)$ and a smaller success probability.}
\label{TNP}
\end{figure}

\begin{figure}
\begin{tabular}{lr}
a)&\\
\multicolumn{2}{c}{\includegraphics*[viewport=3 1 383
296,width=0.9\columnwidth]{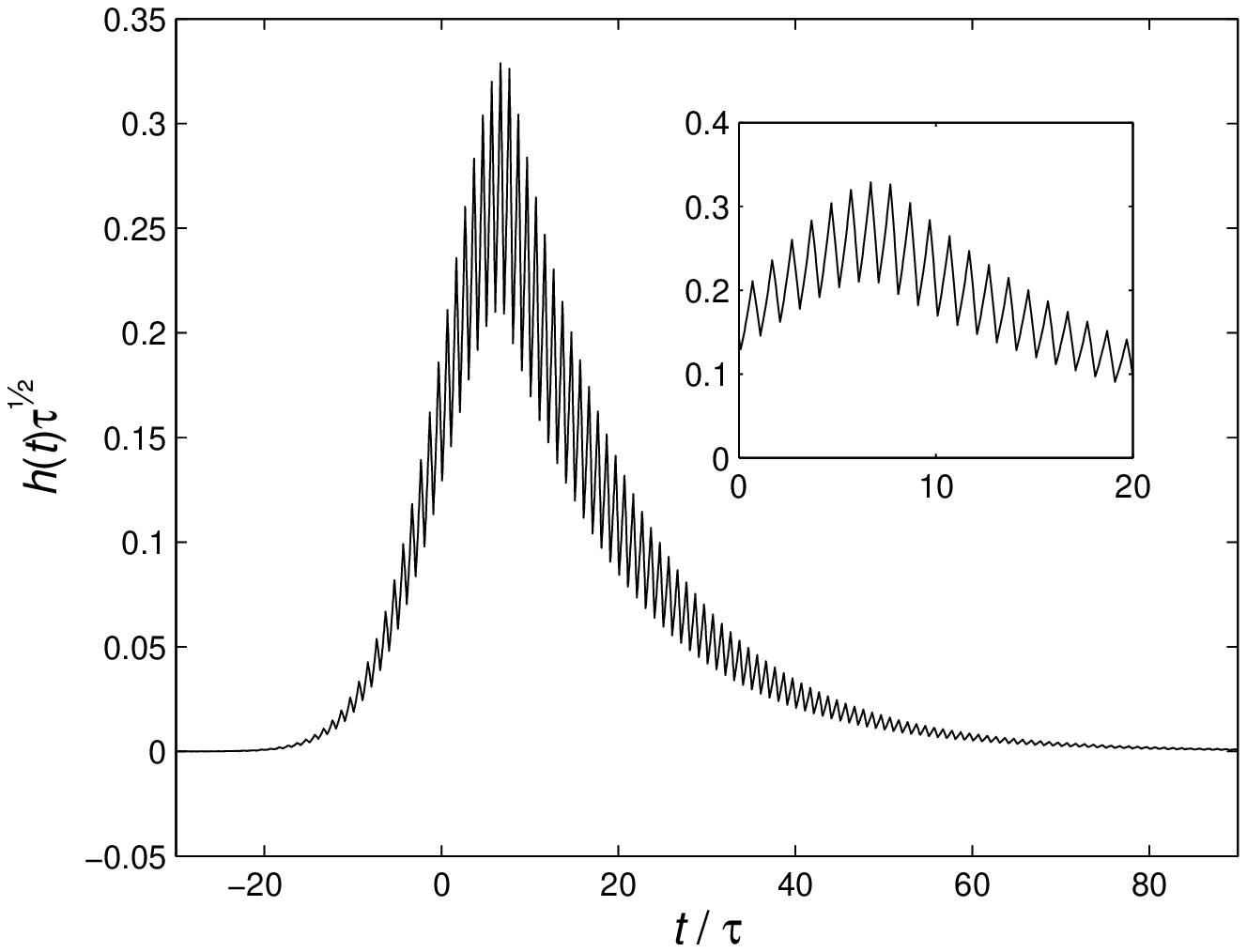}}\\
b)&\multicolumn{1}{l}{c)}\\
\includegraphics*[viewport=3 6 389 296,width=0.45\columnwidth]{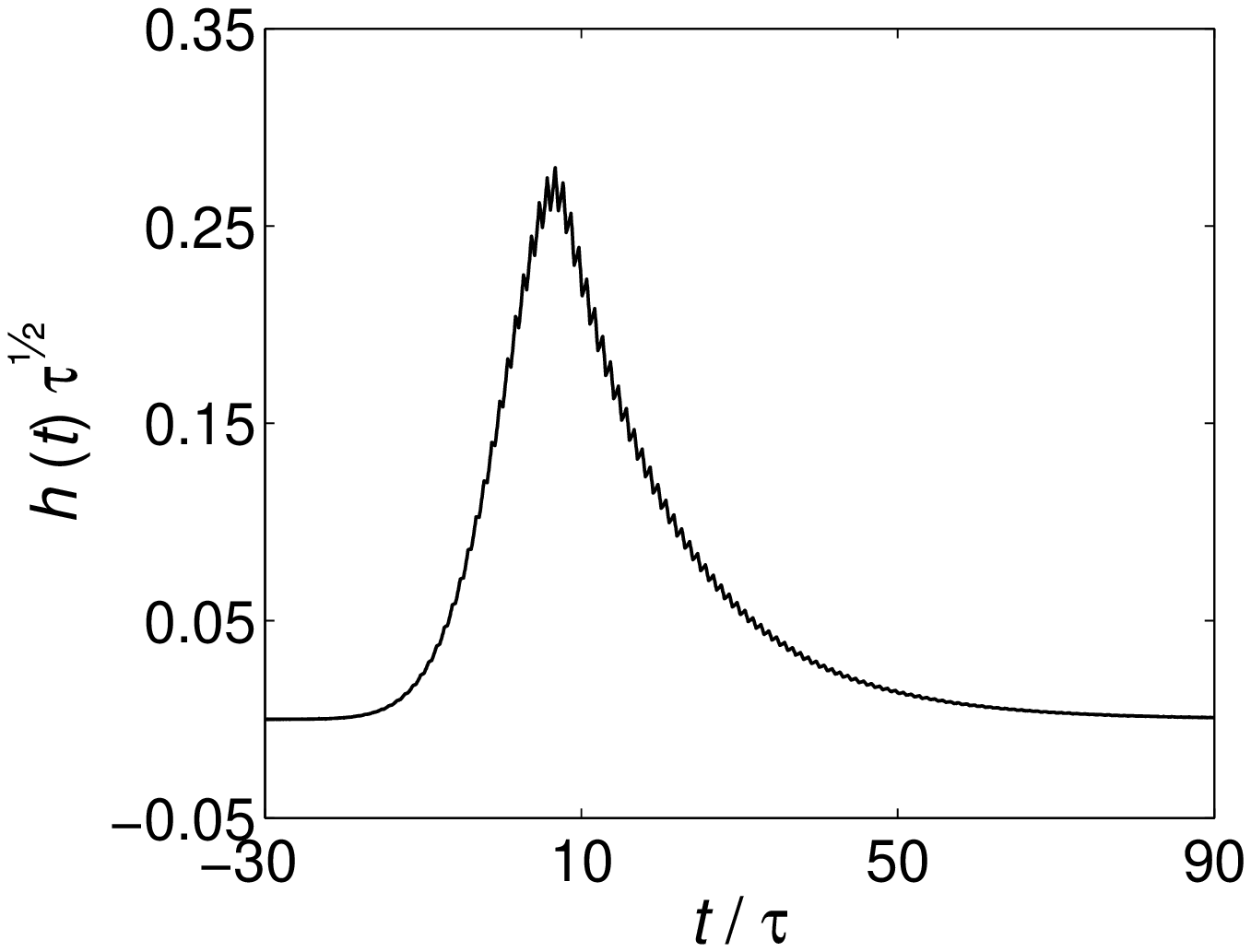}&
\includegraphics*[viewport=3 6 389 296,width=0.45\columnwidth]{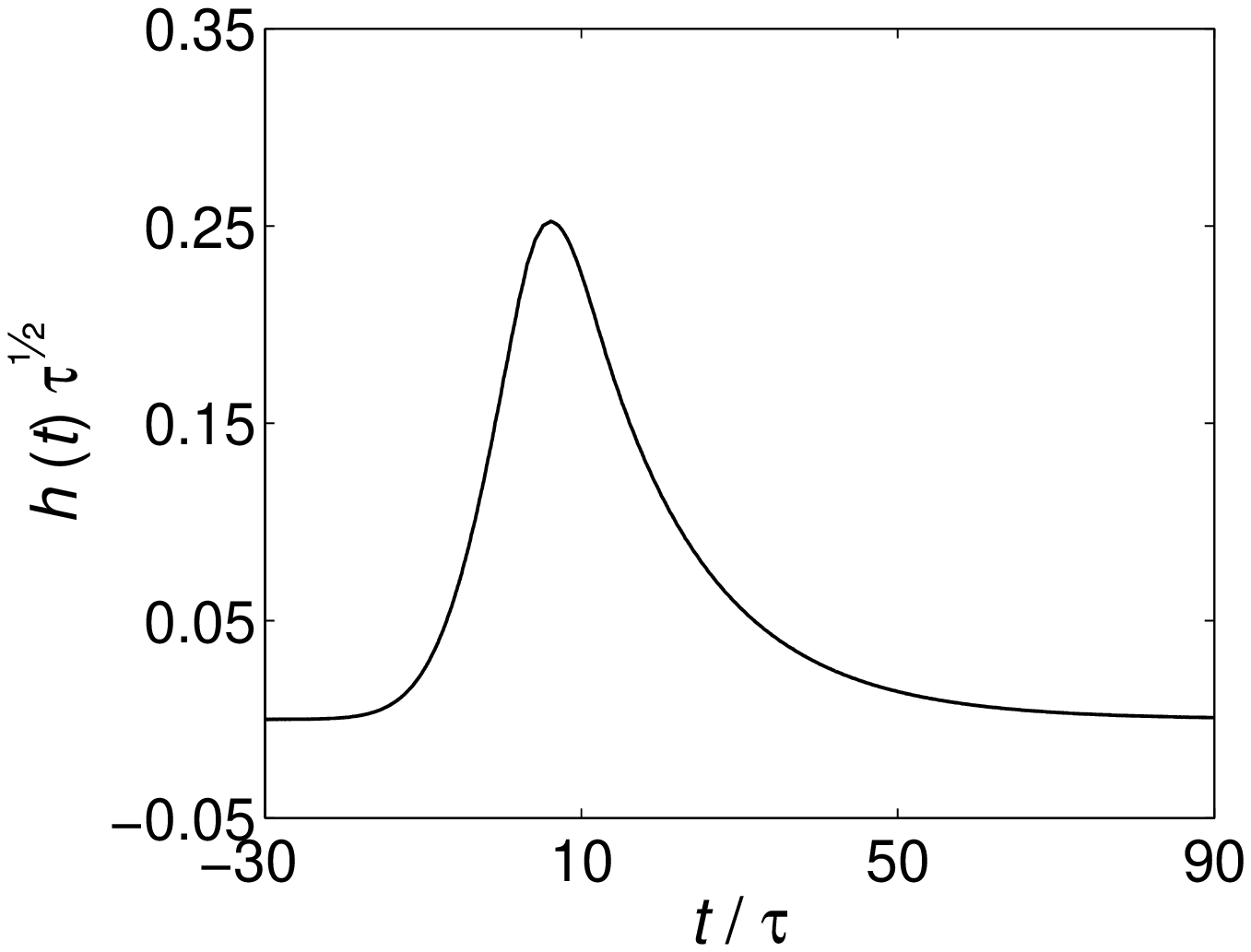}
\end{tabular}
\caption{Mode functions for the modes of the generated state with
the most negative values of $W(0,0)$ for a pulsed pump field with
$T_{\mathrm{p}}/\tau=8$ and $s=0.05$. $t=0$ is chosen to coincide
with the time where $z$ is maximal. a) $T/\tau=0.4$ (the inset is an
enlarged view), b) $T/\tau=2.4$, and c) $T/\tau=10$.}\label{mode}
\end{figure}

\begin{figure}
\begin{center}
\includegraphics*[viewport=4 8 383 296,width=0.95\columnwidth]{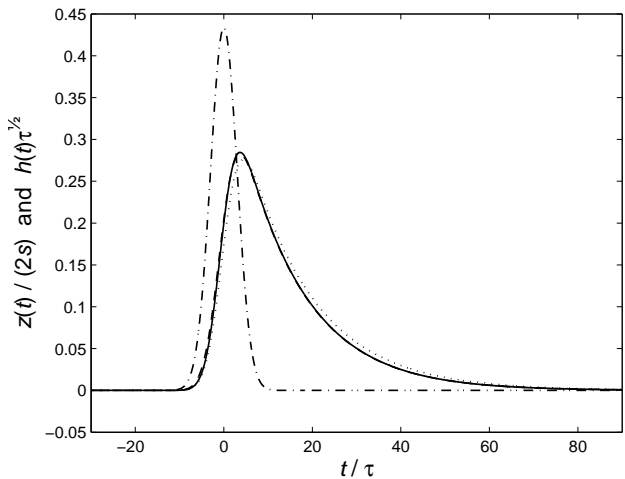}
\end{center}
\caption{Pump pulse (dash-dotted line) and mode function of the mode
of the generated state with the smallest value of $W(0,0)$ (solid
line) for $T_{\textrm{p}}/\tau=3$, $s=0.05$, and $T/\tau=32.4$. The
dotted line is the function in Eq.\ \eqref{th}, while the
approximation \eqref{cah} (dashed line) is almost indistinguishable
from the optimum mode (solid line) in the figure. $W(0,0)=-0.0591$
for the dotted line, $W(0,0)=-0.0610$ for the dashed line, and
$W(0,0)=-0.0611$ for the solid line.} \label{formel}
\end{figure}

Under realistic experimental conditions there will be significant
losses, and these are included in the calculations presented below.
The chosen experimental parameters are given in Table \ref{values},
and for the case of a pulsed OPO we assume that the pump field has a
Gaussian envelope and write $z(t)$ as
\begin{equation}
z(t)=\frac{2s\tau^{1/2}}{\pi^{1/4}
T_{\mathrm{p}}^{1/2}}\exp\left(-\frac{t^2}{2T_{\mathrm{p}}^2}\right).
\end{equation}
Optimizing \eqref{neg} for different values of $T_p$, $s$, and $T$
(or $z$ and $T$ for the case of a time independent pump field) by
optimizing the shape of the mode function of the mode that is to be
stored, we obtain the values of $W(0,0)$ shown in Fig.\ \ref{TNP}a,
and the corresponding success probabilities are given in Fig.\
\ref{TNP}b. Examples of optimized mode functions are plotted in
Figs.\ \ref{mode} and \ref{formel}. We note that the absolute
temporal position of the interval $T$ has been chosen to maximize
the success probability. Also, the values of $z$ for the OPO driven
with a time independent pump field have been chosen in order to
obtain a total flux of photons in the degenerate mode of the output
field from the OPO, which is of order $2\cdot10^6\textrm{ s}^{-1}$
as in Ref.\ \cite{neergaardcat}. This corresponds to a success
probability of $1.9\cdot10^{-4}$ for $T/\tau=10$. The values of $s$
have been adjusted to obtain comparable success probabilities in the
calculations for pulsed pump fields.

Considering the trends in Fig\ \ref{TNP}b, we find that the success
probability increases when $z$ increases and when $s$ increases for
fixed $T_p$ as expected. The success probability is also an
increasing function of $T$, because more terms are included in the
sum in Eq.\ \eqref{P}, when $T$ increases. For a time independent
pump field the increase is linear, because all the $P_i$'s are
equal, while the success probability levels off to a constant value
for a pulsed pump field when $T$ increases beyond the temporal width
of the intensity distribution of the generated output field, which
is roughly of order the width of the intensity distribution of the
pump pulse plus the mean lifetime of a photon in the OPO cavity
($\approx7.6\textrm{ }\tau$) plus the mean lifetime of a photon in
the filter cavity ($\approx0.7\textrm{ }\tau$). The constant value
is thus reached faster for short pulses. It is apparent from the
figure that a decrease in $T_p$ must be accompanied by an increase
in $s$ in order to keep the success probability unchanged. Both of
these changes increase the required peak value of $z$, i.e., a
larger field strength or a larger conversion efficiency in the
crystal is needed.

To explain the trends in Fig.\ \ref{TNP}a we note that $W(0,0)$ of a
state with density matrix $\rho=\sum_{n=0}^\infty\sum_{m=0}^\infty
c_{n,m}|n\rangle\langle m|$ is
$\sum_{i=0}^\infty(c_{2i,2i}-c_{2i+1,2i+1})/\pi$, and thus an
optimization of $W(0,0)$ is equivalent to a maximization of the odd
photon number components. An ideal single mode squeezed vacuum state
is a superposition of even photon number states, and if a photon is
annihilated from the state, these are converted into odd photon
number states. The present experiment deviates from this ideal
situation in two respects. Firstly, the generated state is a
multimode state, since all the generated photon pairs do not belong
to the same mode. The result is that the overlap between the optimal
mode and the modes of the generated photon pairs is only partial,
and this introduces even photon number components into the state of
the optimal mode. The mechanism is particularly severe if photon
pairs are generated in the outer regions of the optimal mode such
that the overlap between the optimal mode and the mode of the photon
pair is significantly smaller than unity, but also significantly
larger than zero. This suggests that a short pump pulse, where the
down conversion process is only turned on in a short period, will
lead to a more negative value of $W(0,0)$, and this is also what is
observed in Fig.\ \ref{TNP}a. The effect is particularly pronounced
for very large $T$. In this limit $W(0,0)$ approaches a constant
value for pulsed pump fields, because $P_i$ is small for trigger
modes far from the center of the intensity distribution of the
generated state and these modes do thus not contribute significantly
to the sums in Eqs.\ \eqref{neg} and \eqref{P}. For a time
independent pump field, on the other hand, $W(0,0)$ approaches the
value of the unconditional state, because the success probability
approaches unity for very large $T$, and all storage attempts are
accepted as successful. The second deviation from ideal behavior is
that losses degrade the odd photon number states into both odd and
even photon number states, and since a photon number state with a
large number of photons is more fragile than a single-photon state,
it is to be expected that a large intensity of down converted
photons will lead to a less negative value of $W(0,0)$. Short pulses
must be intense in order to keep the success probability close to
$10^{-4}$, and this explains why we do not obtain more negative
values of $W(0,0)$ than those presented in Fig.\ \ref{TNP}a if the
pulse duration is decreased below $T_p/\tau=1$ but rather observe
less negative values for very short pulses.

In Fig.\ \ref{TNP} we have included an example of single pass down
conversion for $T_p/\tau=3$, i.e., we have increased $t_1^2$ to
unity without changing the rest of the parameters in Table
\ref{values}. When the OPO cavity is absent, the temporal width of
the light pulse is only broadened in the filter cavity, and this
broadening is small since the mean lifetime of a photon in this
cavity is only $0.7\textrm{ }\tau$. It is thus possible to determine
from the precise APD photo detection time whether the photon pair
that gives rise to the APD detection was generated in the beginning,
in the middle, or in the end of the pulse. For small $T$ this leads
to a negative value of $W(0,0)$ because we know rather precisely
where the second photon in the pair is, if it has not been lost. On
the other hand, if we do not discriminate between early and late
trigger detection events a single optimal compromise will only have
a small overlap with the actual modes, which are well localized in
time. The resulting rapid increase towards positive values in
$W(0,0)$ with increasing $T$ can be avoided by increasing the mean
lifetime of a photon in the filter cavity, corresponding to a
smaller frequency width of the filter, and thus the intensity of the
light transmitted through the filter will decrease, which leads to
an even smaller success probability than in Fig.\ \ref{TNP}b. The
success probability can be increased by increasing $s$, but this
will also result in a less negative value of $W(0,0)$, and from
these arguments it is preferable to apply pulsed fields and the OPO
cavity in the setup to get the most efficient generation of
non-Gaussian states for storage.

The values of $W(0,0)$ presented in Fig.\ \ref{TNP}a are
significantly above the theoretical minimum of $-1/\pi$, and we note
that the results are all obtained for physical parameters, which are
already achieved in experiments. More negative values of $W(0,0)$
can be reached if it is possible to reduce the losses. If, for
instance, we increase the signal channel transmission from $0.7$ to
unity, $W(0,0)$ is decreased to $-0.23$ for $T_p/\tau=3$, $s=0.05$,
and large $T$, while the success probability is unchanged, and if we
further assume zero loss in the cavity and reduce $R$ to $0.01$, we
find $W(0,0)=-0.27$. An increase in the trigger channel transmission
or in the APD detector efficiency will increase the success
probability, and this increase can be transformed into a more
negative value of $W(0,0)$ by decreasing $z$, $s$, $T$, or $R$.

Considering the optimized mode functions for a pulsed pump field in
Fig.\ \ref{mode}, it is apparent that a fast variation in the mode
function is present for small values of $T$, but this variation is
smoothed out for larger $T$. The inset in Fig.\ \ref{mode}a shows
that the period of the fast variation is $\tau$, i.e., the round
trip time in the OPO cavity. A photon generated in one of the time
localized modes in the OPO cavity can only leave the cavity at times
separated by an integer number of $\tau$'s, and thus, in the output
field from the OPO, the two photons in a pair are separated by an
integer number of $\tau$'s. The fast variation in the optimal mode
function can thus be explained from the fact that the mean lifetime
of a photon in the filter cavity is smaller than $\tau$, since this
means that the observed photon in a pair is typically delayed less
than $\tau$ in the filter cavity, and the detection time provides
partial information on the time of generation of the photon pair.
The information is erased as $T$ approaches $\tau$, and this leads
to the steep rise in $W(0,0)$ towards less negative values observed
for all curves to the very left in Fig.\ \ref{TNP}a.

If the duration of the pump pulse is short compared to $\tau$, the
optimal mode function differs qualitatively from those presented in
Fig.\ \ref{mode}. In this case the optimal mode function is a series
of spikes separated by $\tau$, and the width of the spikes is
determined by the width of the pump pulse provided the width of the
pump pulse is also short compared to $T$. In the opposite limit of a
time independent pump field, the optimal mode function for short $T$
is qualitatively a function, which has a maximum close to the time
of the trigger detection and decays exponentially when moving to the
right or left from this maximum. Superimposed on this is a fast
variation with period $\tau$. The mode function is not completely
symmetric around the time of the trigger detection due to the
filtering. For larger values of $T$ the fast variation disappears,
and the maximum becomes more rounded.

In Fig.\ \ref{formel} we show the temporal shape of the pump field
and the optimal mode function for $T/\tau=32.4$ in the same graph,
and it is apparent that the generated output field is both broadened
and delayed due to the nonzero lifetime of a photon in the OPO
cavity as expected. Since $W(0,0)$ and $P$ are both independent of
$T$ for large $T$, and since we have to choose the same mode
function independent of the precise time for the APD detection, the
optimal mode function is determined solely from the generated output
field when $T$ is large, and it seems reasonable that the optimal
mode function just follows the square root of the intensity
distribution of the generated state. Without applying the above
theory, a crude guess for the optimal mode function is
\begin{equation}\label{th}
h(t)\propto\int_{-\infty}^tz(t')\exp\left(-\frac{t_1^2+r_2^2}{2\tau}(t-t')\right)dt',
\end{equation}
and this function, which is also shown in the figure, is actually
quite close to optimal. With the above theory at hand, however, it
is easy to calculate the actual intensity distribution from Eq.\
\eqref{bdb}, and this leads to the mode function
\begin{equation}\label{cah}
h(t)\propto\sqrt{\sum_{n=0}^\infty(r_1t_2)^{2n}
\sinh^2\left(\sum_{k=1}^{n+1}z(t-k\tau)\right)},
\end{equation}
which is even closer to the optimal mode function. Due to its
simplicity, Eq.\ \eqref{cah} can be useful in a storage experiment,
where $T$ must be sufficiently large to obtain a satisfactory
success probability. We note, however, that neither Eq.\ \eqref{th}
nor Eq.\ \eqref{cah} is able to reproduce the optimal mode function
in Fig.\ \ref{mode}a for $T/\tau=0.4$.

\section{Conclusion}

In conclusion, we have presented a multimode description in time
domain of the optical parametric oscillator, which is valid for both
pulsed and continuous-wave pump fields, and we have used the theory
to analyze non-Gaussian states that can be stored in atomic samples.
So far light storage has been demonstrated with Gaussian states
only, and the storage and retrieval of a non-Gaussian state is a
hall mark in the demonstration of atom-light quantum interfaces. Our
analysis suggests that, with similar success probabilities, pulsed
pump fields lead to more negative values of the Wigner function of
the stored state than time independent continuous-wave pump fields.

An essential ingredient in the experiment is the temporal spreading
induced by the OPO and filter cavities, because it separates the two
photons in a generated photon pair temporally by an unknown amount
and ensure that it is impossible to infer the precise position of
the second photon in a pair from the time at which a detection took
place in the APD. Without this spreading we do not obtain large
negative values of $W(0,0)$ for large $T$. After having thus
characterized the system and identified the optimum strategy for
given setup and pulse parameters, one may now move a step further
and try to design filters and pump pulses in order to generate
optimal mode functions which have high non-Gaussian state content
and which are particularly easy to handle in the storage part of an
experiment.

In the present treatment we have chosen to use the negative value
attained by the Wigner function as a measure of the non-classical
character of the state generated, but we note that only minor
changes are required in order to optimize other features as for
example the single-photon or Schr\"odinger kitten state fidelity.

The authors acknowledge discussions with Jonas S.\
Neergaard-Nielsen, Anders S.\ S{\o}rensen, and Eugene S.\ Polzik.

\appendix

\section{Discussion of the applied detector model}

Various detector models are discussed in the literature. Considering
first detection in a single mode, a detector may for instance be a
photon number resolving detector or an on/off detector. The photon
number resolving detector registers the number of photons in the
detected mode, and the detected mode is projected onto the detected
photon number state, while the on/off detector only distinguishes
between the outcomes 'vacuum' and 'not vacuum', and, depending on
the measurement result, the detected mode is projected onto one of
these subspaces. If a continuous beam of light is observed, a third
detector model is often assumed, where a detection event at time $t$
corresponds to an annihilation of a photon in a mode of
infinitesimal temporal width positioned at time $t$. Eq.\ \eqref{Wi}
is derived under the assumption that a detection event is equivalent
to application of the annihilation operator of the observed mode to
the state of the system, but since the flux of trigger photons is
small in the experiment considered in Sec.\ IV and the possibility
to have two or more photons in a trigger mode can be neglected, all
three detector models lead to practically identical results.

A real detector consists of a medium, which absorbs photons, and
typically several different microscopic transitions in the medium
are allowed. The detector is then able to detect photons in more
than a single mode. In an APD a photon absorption is accompanied by
an excitation of an electron from a bound state to a free or a solid
state conduction band state, the signal is amplified, and the
resulting photo current is measured. The information concerning the
precise microscopic transition is lost in the amplification process,
and, when we condition on the macroscopic detection event, we must
average over all modes that could have led to the observed detection
(see \cite{rohde}). Since the filter in front of the detector
selects a narrow band of frequencies, we assume that the detector is
equally sensitive to all frequencies transmitted by the filter,
i.e., the detector bandwidth is assumed to be infinitely broad
compared to the filter bandwidth, and all modes are observed by the
detector. In this case we should sum over a complete set of modes on
the interval $T$ in Eq.\ \eqref{neg}. The obtained values of
$W(0,0)$ and $P$ are independent of the choice of basis which may be
seen from the following argument. Assume that the mode functions
$f_i$ and the mode functions $g_i$ both constitute a complete basis
on the time interval $T$. $f_i$ and $g_i$ are related by a unitary
transformation $f_i=\sum_jU_{ij}g_j$, which implies that the
corresponding mode annihilation operators $\hat{a}^f_i=\int
f_i^*(t)\hat{a}(t)dt$ and $\hat{a}^g_i=\int g_i^*(t)\hat{a}(t)dt$
are related by $\hat{a}^f_i=\sum_jU_{ij}^*\hat{a}^g_j$. If a photon
is annihilated in the mode $\hat{a}^f_i$, the density matrix is
transformed according to
$\rho\rightarrow\hat{a}^f_i\rho(\hat{a}^f_i)^\dag/
\mathrm{Tr}(\hat{a}^f_i\rho(\hat{a}^f_i)^\dag)$, where Tr denotes
the trace. To perform the averaging we multiply by the probability
$\mathrm{Tr}(\hat{a}^f_i\rho(\hat{a}^f_i)^\dag)$ to annihilate a
photon in the mode $\hat{a}^f_i$, and sum over all $i$. After
normalization this leads to
\begin{equation}
\rho\rightarrow\frac{\sum_i\hat{a}^f_i\rho(\hat{a}^f_i)^\dag}
{\textrm{Tr}(\sum_i\hat{a}^f_i\rho(\hat{a}^f_i)^\dag)}.
\end{equation}
Since $\sum_i\hat{a}^f_i\rho(\hat{a}^f_i)^\dag
=\sum_i\sum_jU_{ij}^*\hat{a}^g_j\rho\sum_kU_{ik}(\hat{a}^g_k)^\dag
=\sum_j\sum_k\hat{a}^g_j\rho(\hat{a}^g_k)^\dag\delta_{jk}
=\sum_i\hat{a}^g_i\rho(\hat{a}^g_i)^\dag$, the transformed density
matrix is independent of the choice of basis, and the Wigner
function is then also independent of the choice of basis. It is thus
not necessary to use the modes, in which the detections actually
take place, and for mathematical convenience, we have chosen to sum
over time localized modes in Eq.\ \eqref{neg}.


\begin{thebibliography}{99}
\bibitem{collett} M. J. Collett and C. W. Gardiner, Phys. Rev. A
{\bf30}, 1386 (1984).
\bibitem{drummond} P. D. Drummond and M. D. Reid, Phys. Rev. A
{\bf41}, 3930 (1990).
\bibitem{lu} Y. J. Lu and Z. Y. Ou, Phys. Rev. A {\bf62}, 033804
(2000).
\bibitem{adamyan} H. H. Adamyan and G. Yu. Kryuchkyan, Phys. Rev. A
{\bf74}, 023810 (2006).
\bibitem{bennink} R. S. Bennink and R. W. Boyd, Phys. Rev. A
{\bf66}, 053815 (2002).
\bibitem{wasilewski} W. Wasilewski, A. I. Lvovsky, K. Banaszek, and
C. Radzewicz, Phys. Rev. A {\bf73}, 063819 (2006).
\bibitem{petersen} V. Petersen, L. B. Madsen, and K. M{\o}lmer,
Phys. Rev. A {\bf72}, 053812 (2005).
\bibitem{ourjoumtsevcat} A. Ourjoumtsev, R. Tualle-Brouri, J. Laurat,
and P. Grangier, Science {\bf 312}, 83 (2006).
\bibitem{neergaardcat} J. S. Neergaard-Nielsen, B. M. Nielsen,
C. Hettich, K. M{\o}lmer, and E. S. Polzik, Phys. Rev. Lett.
{\bf97}, 083604 (2006).
\bibitem{wakui} K. Wakui, H. Takahashi, A. Furusawa, and M. Sasaki,
Opt. Express {\bf 15}, 3568 (2007).
\bibitem{nielsen1} A. E. B. Nielsen and K. M{\o}lmer, Phys. Rev. A {\bf 75},
023806 (2007).
\bibitem{andre} A. Andr\'e, M. Bajcsy, A. S. Zibrov, and M. D. Lukin,
Phys. Rev. Lett. {\bf 94}, 063902 (2005).
\bibitem{eit} S. E. Harris, Phys. Today \textbf{50}, No. 7, 36 (1997).
\bibitem{storeraman} A. E. Kozhekin, K. M{\o}lmer, and E. Polzik,
Phys. Rev. A \textbf{62}, 033809 (2000).
\bibitem{storerot} B. Julsgaard, J. Sherson, J. I. Cirac, J. Fiur\'{a}\v{s}ek,
and E. S. Polzik, Nature \textbf{432}, 482 (2004).
\bibitem{sherson} J. Sherson, A. S. S{\o}rensen, J. Fiur\'a\v sek, K.
M{\o}lmer, and E. S. Polzik, Phys. Rev. A {\bf74}, 011802(R) (2006).
\bibitem{storegen} I. Novikova, A. V. Gorshkov, D. F. Phillips, A. S.
S\o rensen, M. D. Lukin, and R. L. Walsworth, Phys. Rev. Lett.
\textbf{98}, 243602 (2007).
\bibitem{blow} K. J. Blow, R. Loudon, S. J. D. Phoenix, and T. J. Shepherd, Phys.
Rev. A {\bf42}, 4102 (1990).
\bibitem{neergaard1} J. S. Neergaard-Nielsen, B. M. Nielsen, H.
Takahashi, A. I. Vistnes, and E. S. Polzik, Opt. Express {\bf15},
7940 (2007).
\bibitem{rohde} P. P. Rohde and T. C. Ralph,
J. Mod. Opt. \textbf{53}, 1589 (2006).
\end{thebibliography}
\end{document}